\documentstyle[12pt]{article}
\ifx\epsffile\undefined
\newlength{\epsfysize}
\def\epsffile#1#2#3#4]#5{}
\else
\fi

\def\dr{\mbox{$\overline{\rm DR}\ $}}
\def\ds{\mbox{$\scriptscriptstyle\overline{\rm DR}\ $}}
\def\roughly#1{\raise.3ex\hbox{$#1$\kern-.75em\lower1ex\hbox{$\sim$}}}

\begin{document}
\begin{titlepage}
\begin{center}
			   \hfill JHU-TIPAC-940010\\
		           \hfill hep-ph/9407202\\
\vskip .9 in
{\large \bf Weak Scale Threshold Corrections in Supersymmetric Models
\footnote{Talk presented at the SUSY-94 Conference, Ann Arbor, Michigan,
May 14--17, 1994.}}
\vskip .3 in
           \vskip 0.5 cm
      {\bf Damien Pierce}\\
\vskip.2in
      {\it Department of Physics and Astronomy\\
          The Johns Hopkins University\\
         Baltimore, Maryland\ \ 21218\\}
          \vskip 0.5 cm
\end{center}
\vskip 0.4 in
\begin{abstract}
I discuss the weak scale threshold corrections in supersymmetric models.
I describe the ``match and run" approximation to the threshold
corrections and compare with the exact one-loop results. With
explicit examples I show that in cases without large hierarchies in the mass
spectra the ``match and run" approximation can lead to order $\cal{O}$(1)
errors in the determination of the threshold corrections. I demonstrate
how to obtain the threshold-corrected Yukawa coupling from the
fermion pole mass. I present corrections to the top quark and
squark/slepton masses as a
function of the GUT scale parameters $m_0$ and $m_{1/2}$ and show
that the gauge/Higgs sector corrections to the top quark mass are small
while the gluino correction can be larger than the
well known gluon correction.
\end{abstract}
\end{titlepage}
\renewcommand{\thepage}{\arabic{page}}
\setcounter{page}{1}

\section{Introduction}
I will contrast two methods of accounting for weak scale threshold
corrections. Although the discussion will be quite general I will
concentrate on obtaining the physical spectrum of particle masses from a
set of \dr running parameters in the context of supersymmetric GUT
models.

In a supersymmetric GUT model with supergravity boundary conditions, we
have as inputs at the GUT scale the universal scalar mass $m_0$, the
gaugino mass $m_{1/2}$, the A-term $A_0$, the gauge coupling
$\alpha_{\rm GUT}$ and the Yukawa couplings $\lambda_i$.  We then run
these parameters down to low energies using the two-loop \dr renormalization
group equations\cite{rge}. The weak scale threshold corrections to the masses
take us from the running \dr masses, evaluated at an arbitrary \dr
scale $\mu$, to the physical mass. Hence we have the correspondence
$$ {\rm Weak\ scale\ threshold\ corrections: }\qquad
m_{\ds}\!(\mu)\ \ \longleftrightarrow\ \ m_{\rm pole}\ .$$

In the next section I compare the threshold corrections
obtained using the ``match and run" method with the exact one-loop
results. In Sec.3 I discuss how to obtain the threshold corrected
Yukawa couplings from the physical quark masses. In Sec.4 I present some
results, and I briefly summarize.

\section{Comparison}
In this section I first describe the ``match and run" approximation
and then explicitly compare with the exact one-loop results.
The ``match and run" procedure is  often used in the literature
for approximating the weak scale threshold corrections.
The advantage of this method, and perhaps the main
reason for its ubiquity, is that one may approximate
the threshold corrections with only a
knowledge of the RGE's. The procedure is based
on effective field theory and the decoupling theorem. Solving the \dr
renormalization group equation for a given
mass parameter
$${dm\over dt} = {\beta\over16\pi^2}m,\qquad t=\ln\mu^2$$
with the boundary condition $m(M_{\rm GUT}) = m_0$, we obtain
the \dr running mass $m(\mu)$. In solving this equation with the
one-loop $\beta$-function we sum the leading logarithms of the type
$\sum_n \left({\beta/16\pi^2}\right)^n\ln^n\left({M^2_{\rm
GUT}/\mu^2}\right)$. The one-loop $\beta$-function, with the form
$\beta = \sum_ic_{ij}g_j^2$, is (for example) a sum of
(gauge or Yukawa) coupling constants squared multiplied by quadratic
Casimir coefficients $c_{ij}$, and the contributions in the sum
correspond to physical particles circulating in a one-loop diagram.

Consider a particle of mass $m$ which
receives contributions to its $\beta$-function from all the various
particles in the MSSM.
At scales larger than the heaviest particle in the spectrum (e.g.
the squarks) the particle's \dr mass evolves according
to the full MSSM RGE. As we decrease $\mu$ we eventually
encounter the scale of the squark masses. At this point we stop the
RG evolution and construct a new effective theory in which the squarks
are integrated out. At scales below the squark mass the squarks are not
active degrees of freedom; they do not circulate in the loops. Hence,
in the effective theory at scales below the squark mass we do not
include the squark contributions to the $\beta$-functions.
Also we match the two theories at the scale $\mu=m_{\tilde{q}}$.
Hence we set $m(m_{sq}^-) = m(m_{sq}^+)$, and continue
\begin{figure}[t]
\epsfysize=2.0in
\begin{center}
\epsffile[-95 220 610 475]{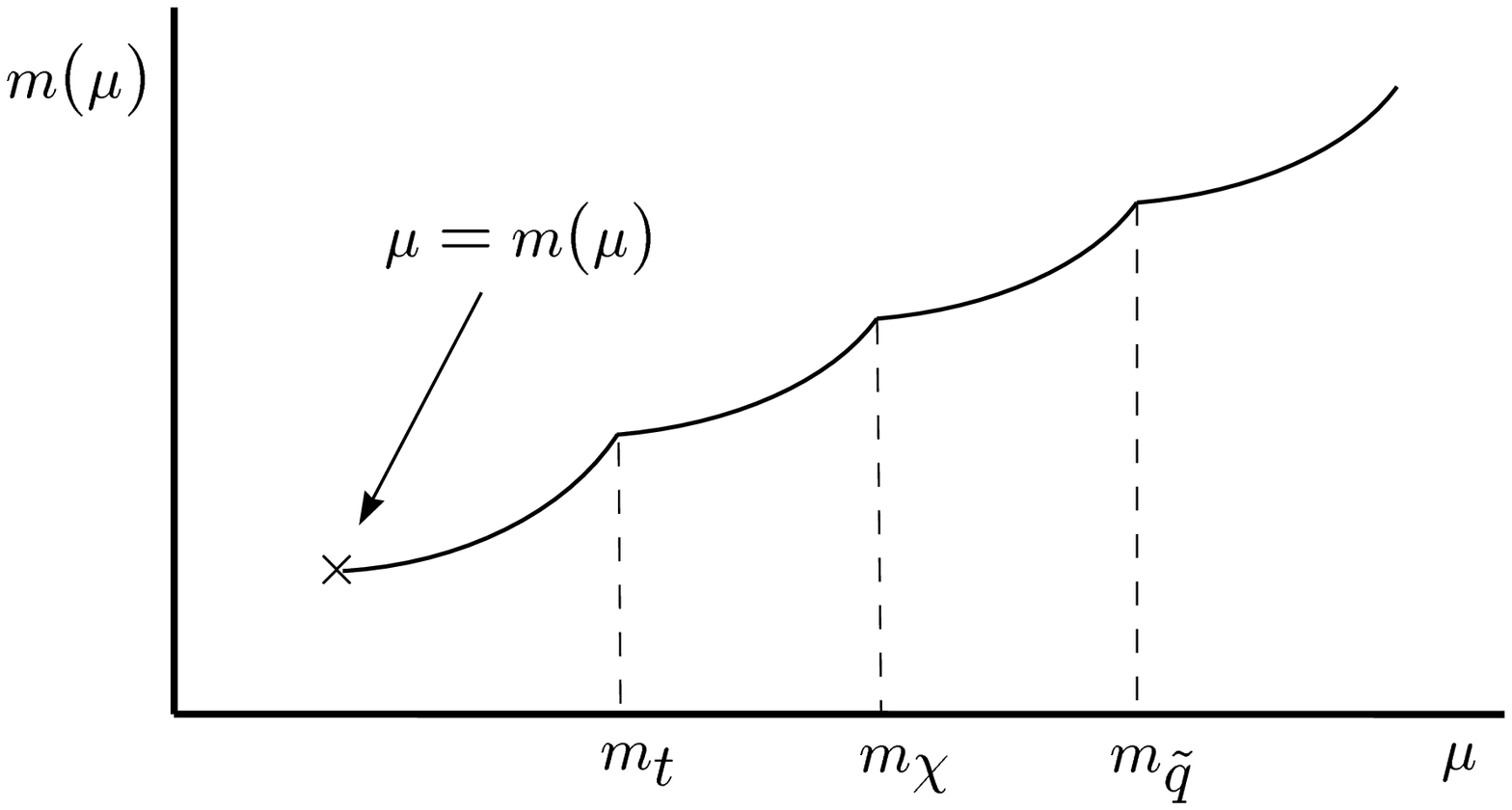}
\parbox{5.5in}{
\caption[]{\small The \dr mass $m(\mu)$ vs. the \dr scale $\mu$.
}}
\end{center}
\end{figure}
the RG evolution of the parameter $m$ with the new $\beta$-function until
we reach the next heaviest particle that contributes to the RGE
(e.g. the $\chi$'s). At this point we subtract the chargino
and neutralino
contributions to the $\beta$-functions, require continuity of the
parameters, and continue. Likewise we decouple the top quark, the
$Z$-boson, and so on. The RG evolution is then terminated
when we reach the scale $\mu=m(\mu)$. This quantity $m(m)$ is then the
approximation to the physical mass in the ``match and run" program.
This procedure is illustrated in Fig.1.

We now show how the ``match and run" procedure compares
with the exact one-loop result. Consider the threshold correction for a
particle of mass $m$ due to a particle of mass $M$. First we consider
the case $M > m$. As can be seen from Fig.2,
\begin{figure}[htb]
\epsfysize=3in
\epsffile[10 445 610 720]{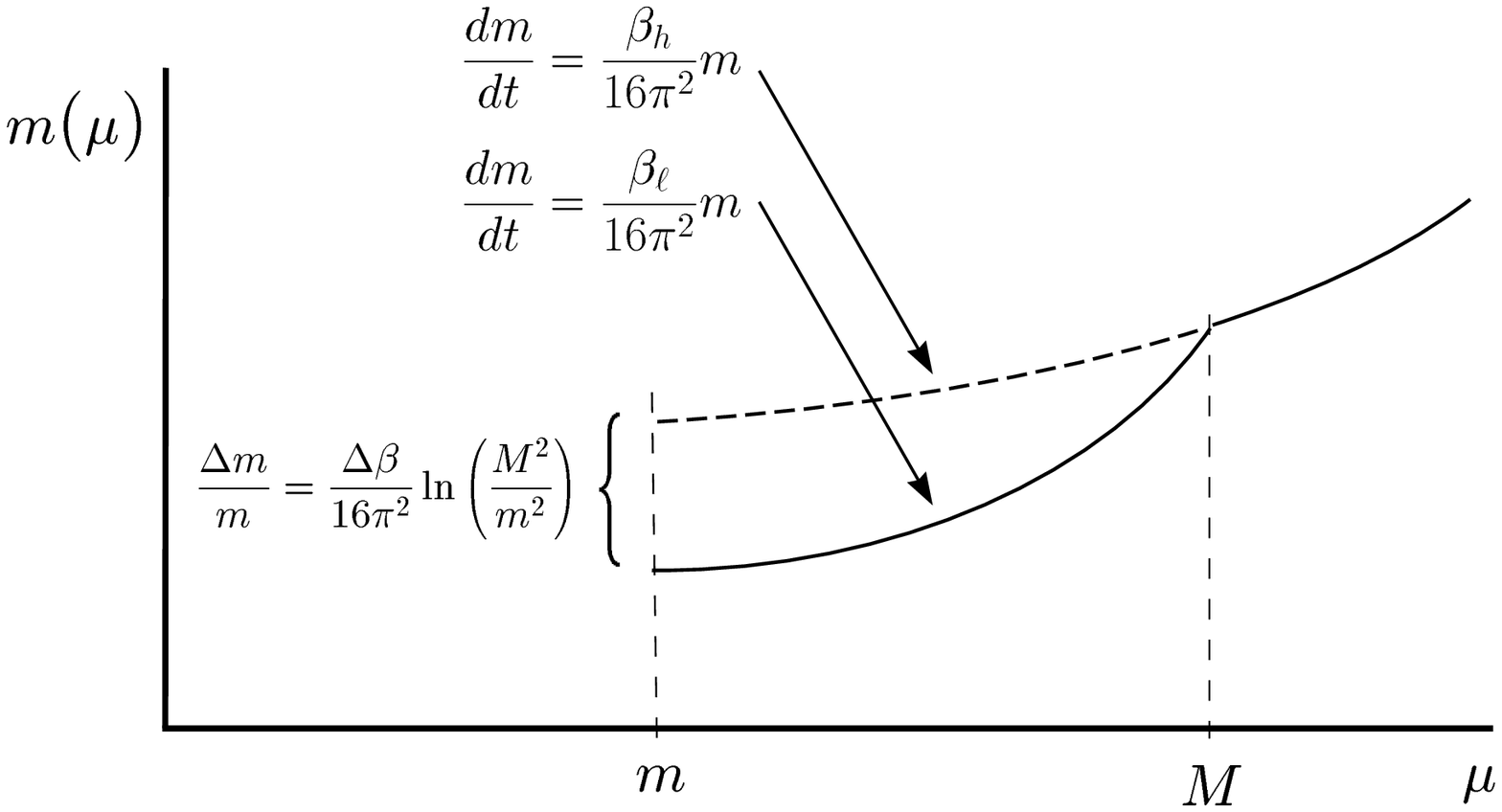}
\begin{center}
\parbox{5.5in}{
\caption[]{\small The \dr mass $m(\mu)$ vs. the \dr scale $\mu$.
The correction is shown in the case that one particle of
mass $M>m$ is decoupled.}}
\end{center}
\end{figure}
decoupling the particle of mass $M$ yields the correction
$${\Delta m\over m} = {\Delta\beta\over16\pi^2}\ln\left(
{M^2\over m^2}\right),\qquad (M>m)\ ,$$
where $\Delta\beta$ = $\beta_h-\beta_\ell$. $\beta_h(\beta_\ell)$ is the
$\beta$-function including (not including) the contribution from the
heavy particle of mass $M$. For the case $M\leq m$,
the ``match and run" procedure gives the correction
$${\Delta m\over m} = 0,\qquad (M\leq m)\ .$$

Now we consider the exact one-loop result. Upon evaluating
the diagram of Fig.3 we find
\begin{figure}[b]
\epsfysize=1.3in
\epsffile[-25 575 610 730]{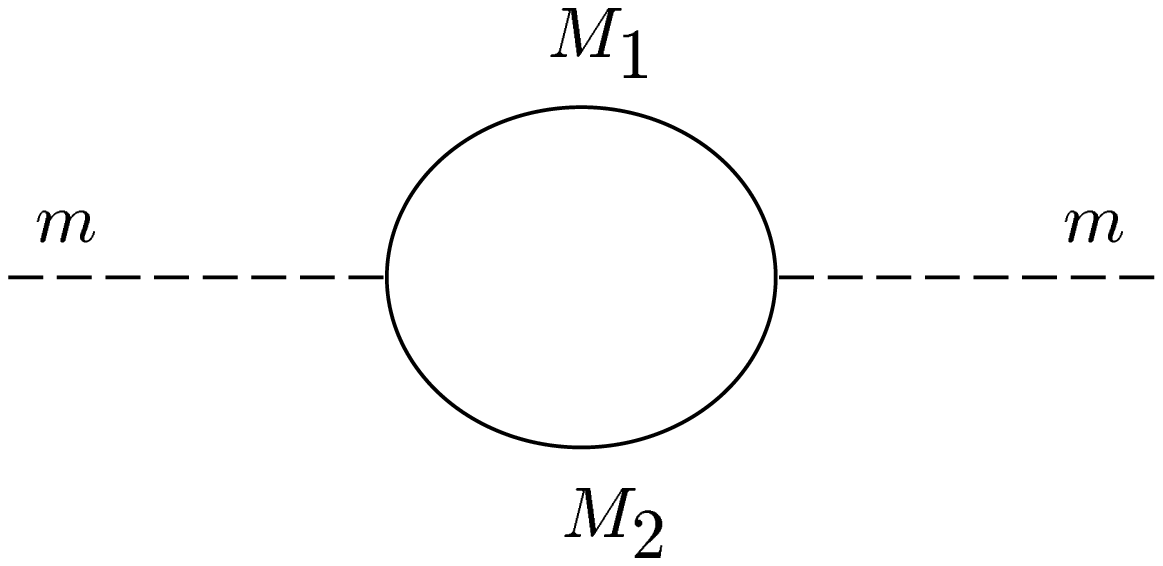}
\begin{center}
\parbox{5.5in}{
\caption[]{\small
The one-loop diagram yielding the threshold correction for a
particle of mass $m$ due to particles with masses $M_1$ and $M_2$.
}}
\end{center}
\end{figure}

$${\Delta m\over m}
= {\Delta\beta\over16\pi^2}\int_0^1\ dx\ \ln\left(
{\left|(1-x)\,M_1^2 + x\,M_2^2 - x(1-x)\,m^2\right|\over\mu^2}\right)\ .$$
In case of $M_1=M_2 >m$ (and setting $\mu=m$) we find the
correction
$${\Delta m\over m} = {\Delta\beta\over16\pi^2}\biggl[\ln\left(
{M^2\over m^2}\right)+ {\rm finite}\biggr],\qquad (M>m)\ ,$$
In the case $M_1, M_2 \leq m$ we find
$${\Delta m\over m} = {\Delta\beta\over16\pi^2}\biggl[0
+ {\rm finite}\biggr],\qquad (M\leq m)\ ,$$
where the `0' signifies no logarithmic correction.

The ``match and run" procedure leads to good approximations to
the pole masses when the mass under consideration $m$ is much smaller than
the masses of the decoupled particles $M_i$. In the limit $M_i^2\gg m^2$
the large logarithmic corrections proportional to $\log\left({M_i^2/
m^2}\right)$ are correctly taken into account.

On the other hand, at each threshold there are
finite corrections which are entirely missed in the ``match and run"
framework. These finite (i.e. not logarithmically enhanced)
corrections can be as large as the logarithmic corrections when, for
example,
all of the particle masses are of the same order of magnitude. In fact,
in supersymmetric GUT models with universal boundary conditions it is
not uncommon that the entire supersymmetric spectrum is of order $M_Z$.
In such a case we can expect the finite corrections to be as large as
the logarithms. Hence, the error in evaluating the threshold corrections
in the ``match and run" procedure can be ${\cal O}(1)$.

Since the exact one-loop threshold functions for all of the
particles in the
MSSM have been calculated\cite{corrs} both the logarithmic and finite
corrections can be consistently incorporated, and this leads to precise
results. Furthermore,
with the threshold functions in hand, the \dr scale has no significance.
In the ``match and run" procedure it was important to stop the running
of a mass at the scale equal to the mass. However, when using the
one-loop threshold functions the \dr parameters can be
evaluated at any scale of order the electroweak scale and
no decoupling in the RG evolution is necessary.
Hence, we can simply run all the \dr parameters down to the scale $M_Z$
using the original set of RG equations, and then add the threshold
corrections to obtain the pole masses.

Of course when using the two-loop RG equations it is important to
include the threshold corrections correctly. The corrections due
to the two-loop RG running are expected to be numerically
of the same order as the one-loop
threshold corrections and as we have stated above the ``match and run"
procedure can lead to ${\cal O}(1)$ errors in the determination of the
threshold corrections. Hence we emphasize that the following
go together:
$$\rm \biggl\{ \ two\mbox{-}loop~~RGE\mbox{'}s,~~~one
\mbox{-}loop~~threshold~~functions\ \biggr\}\ .$$

\section{Examples}
In this section I list a few examples of finite threshold corrections,
then I show some examples comparing the ``run and match" approximations
with the exact one-loop corrections.

The correction to the top quark mass due to the gluon loop
is well known\footnote{Actually this correction is more
commonly seen as $4\alpha_s/(3\pi)$ which is the result in the
$\overline{\rm MS}$ scheme.}
$$m_t^{\rm pole} = m_t(m_t)\left(1 + {5\alpha_s\over3\pi}\right)\ ,$$
where $m_t(m_t)$ signifies the running \dr mass evaluated at the scale
$m_t$. The left hand side of the above equation is not really the pole
mass, as the top quark mass receives many other corrections.
The next most important is the gluino/squark correction
$${\Delta m_t^{\tilde{g}\tilde{q}}\over m_t} =  -{\alpha_s\over3\pi}\Biggl\{
{\rm Re}\left[B_1(m_t,m_{\tilde{g}},m_{\tilde{t}_1})
+ B_1(m_t,m_{\tilde{g}},m_{\tilde{t}_2}) \right] $$$$
\qquad\qquad\qquad- {2m_{\tilde{g}}\left(A_t+\mu\cot\beta\right)\over
m_{\tilde{t}_1}^2 - m_{\tilde{t}_2}^2}
\,{\rm Re}\left[ B_0(m_t,m_{\tilde{g}},m_{\tilde{t}_1})-
B_0(m_t,m_{\tilde{g}},m_{\tilde{t}_2}) \right]\Biggr\}$$
where $B_0$ and $B_1$ are the two point functions
$$B_n(p;m_1,m_2) = -\int_0^1dx\,x^n\,\ln\left(
{(1-x)\,m_1^2 + x\,m_2^2 - x(1-x)\,p^2\over
\mu^2}\right)\ .$$
This correction for $m_{\tilde{g}}>m_t$ and/or $m_{\tilde{t}}>m_t$
is actually a logarithmic correction plus a finite correction.
For the top quark the gluon correction is about 6\%. As we show in the last
section, the gluino/squark correction is typically of the same order
as the gluon correction.

For the gluino mass we have the finite correction due to the gluon loop
$$m_{\tilde{g}}^{\rm pole} = M_3(M_3)\left(1+{15\alpha_s\over
4\pi}\right)$$
where $M_3(M_3)$ is the \dr gluino mass parameter $M_3$ evaluated at
$M_3$. Here again the left hand side is not really the pole mass, as
there are finite corrections (and potentially large
logarithmic corrections)
from the quark/squark loops as well\cite{Martin,Pierce}.

Another example of an important finite correction arises for the
bottom quark mass, where the second line in the gluino/squark
correction to the top quark mass shown above yields in the case
of the bottom quark
$${\Delta m_b\over m_b} \sim -{2\alpha_s\over 3\pi} {\mu
m_{\tilde{g}}\over m_{\tilde{b}}^2} \tan\beta\qquad
({\rm large}\ \ \tan\beta)\ ,$$ and this correction, which is
entirely missed in the ``match and run" procedure, can be as large as
50\% (see S. Pokorski's talk in the SUSY-94 conference proceedings).

Next I compare the exact and ``run and match" results
quantitatively. In the following figures I show the gluino mass, first in
the case where the squark masses are much larger than the gluino mass,
then in the case where the squarks and gluino are approximately
degenerate. As we expect, the ``run and match" approximation is a better
approximation in the first case than in the second.

In Fig. 4(a) I plot the gluino mass versus the \dr
\begin{figure}[tb]
\epsfysize=3in
\begin{center}
\epsffile[40 510 610 750]{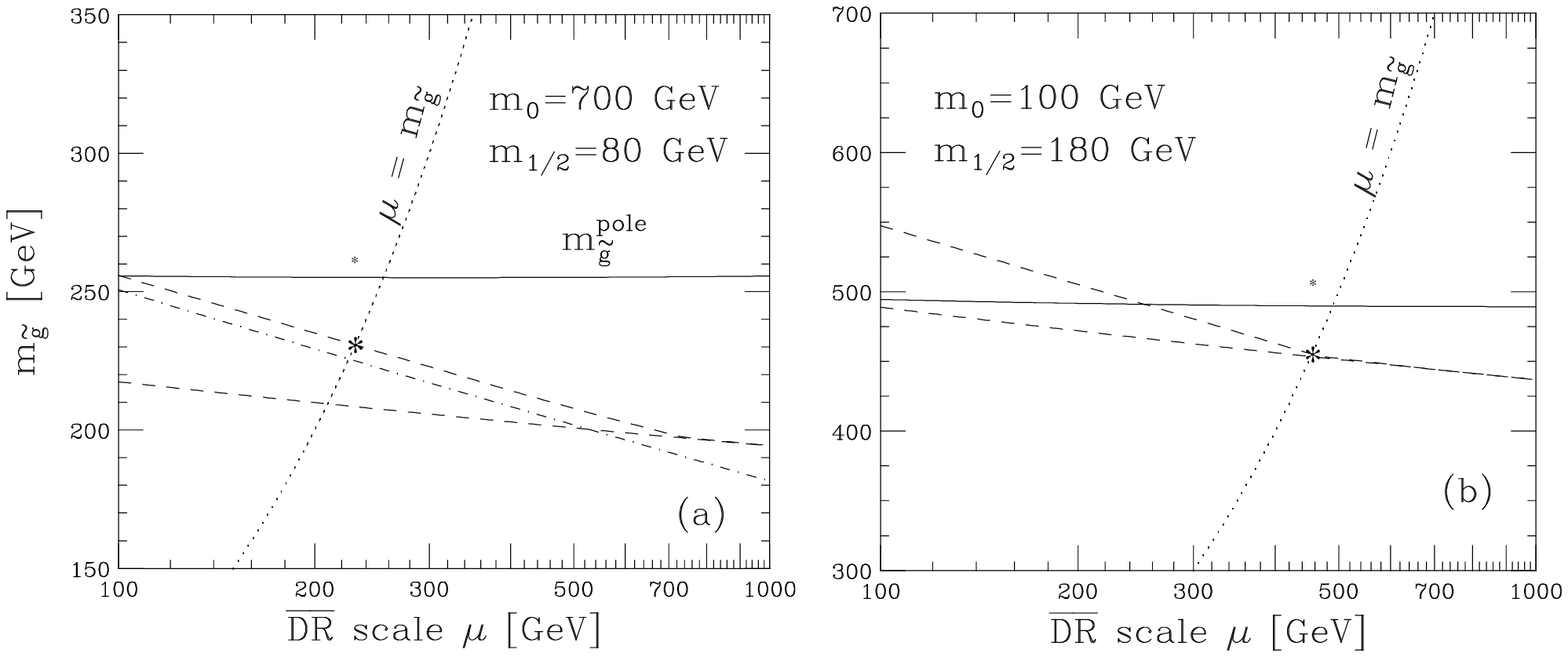}
\parbox{5.5in}{
\caption[]{\small The gluino mass vs. the \dr scale $\mu$. The dashed
lines show the running mass with and without decoupling the squarks.
The solid line indicates the pole mass, and the horizontal dashed line
shows the points $\mu=m_{\tilde{g}}$. In Fig.(a) The dash-dotted line
indicates the exact one-loop squark correction. The asterisks indicate
$m_{\tilde{g}}(m_{\tilde{g}})$ before and after adding the gluon
correction.
}}
\end{center}
\end{figure}
scale $\mu$ setting $m_0$=700 GeV and $m_{1/2}=$80 GeV.
The dashed lines show the \dr running gluino masses.
The lower dashed line shows the undecoupled \dr mass, and the
upper dashed line shows the running gluino mass in which the
squarks are decoupled.
The squarks are decoupled at around 750 GeV in this case. Parallel to
this line and just below it is a dash-dotted line which indicates the
exact one-loop result obtained when taking into account the squark
corrections.
The difference between these two lines indicates the finite correction.
The fact that these two lines are straight and parallel indicates that
the exact and ``match and run" corrections both account for the same
leading logarithm. If we add to the dot-dashed line the exact one-loop
gluon correction we obtain the
physical mass shown as a solid horizontal line, at 255 GeV. The fact
that it is independent of $\mu$ shows that we can evaluate the physical
mass at any
\dr scale in the vicinity of the electroweak scale. This physical mass
is not exactly independent of $\mu$ because solving the
RG equations includes all orders in perturbation theory; the first order
part of the RG result is exactly cancelled by the one-loop threshold
function, and the higher order parts of the running are small when
considering scales in the range $M_Z$ to 1 TeV.

The points $\mu=m_{\tilde{g}}$ are also indicated in Fig. 4(a) as a
dotted line. In
the ``match and run" procedure one would follow the decoupled running
mass line from large scales until it intersects the line
$\mu=m_{\tilde{g}}$ (indicated by large asterisks in the figures)
and this mass $m_{\tilde{g}}(m_{\tilde{g}})$
(230 GeV) would be the approximation to the physical mass. Of course the
finite gluon correction could be included, which yields 263 GeV
(shown as a small asterisk).
In this case, because the squarks are much heavier than the gluino
the logarithmic squark correction is larger than the finite squark
correction. The threshold correction due to the squarks is
approximated by the ``match and run" leading logarithm. At the scale
$\mu=m_{\tilde{g}}$ the ``match and run" correction is 22 GeV and the
exact result is 14 GeV. Even in this case in which
$m_0 \gg m_{1/2}$ the approximation is not so good; this indicates
that the ``large logarithm" $\log(m_{\tilde{q}}^2/m_{\tilde{g}}^2)$
with $m_{\tilde{q}}=750$ GeV and $m_{\tilde{g}}=250$ GeV cannot really
be considered large.

Of course if the squark masses are lighter than or equal to the gluino
mass, the correction due to the squarks is 0 in the ``match and run"
scheme. This is illustrated in Fig. 4(b), where I show again the running
(squark decoupled and undecoupled) \dr gluino masses and the pole mass
as well as the line $\mu=m_{\tilde{g}}$ in the case $m_0$=100 GeV,
$m_{1/2}=180$ GeV. The ``match and run"
mass of 456 GeV underestimates by 8\% the pole mass of 492 GeV.
If the finite gluon contribution is added to the ``match and run"
mass, the result overestimates the pole mass by 4\%.

In Fig.5 I show similar results for the top quark mass. Here the
\begin{figure}[htb]
\epsfysize=3in
\begin{center}
\epsffile[-68 380 400 740]{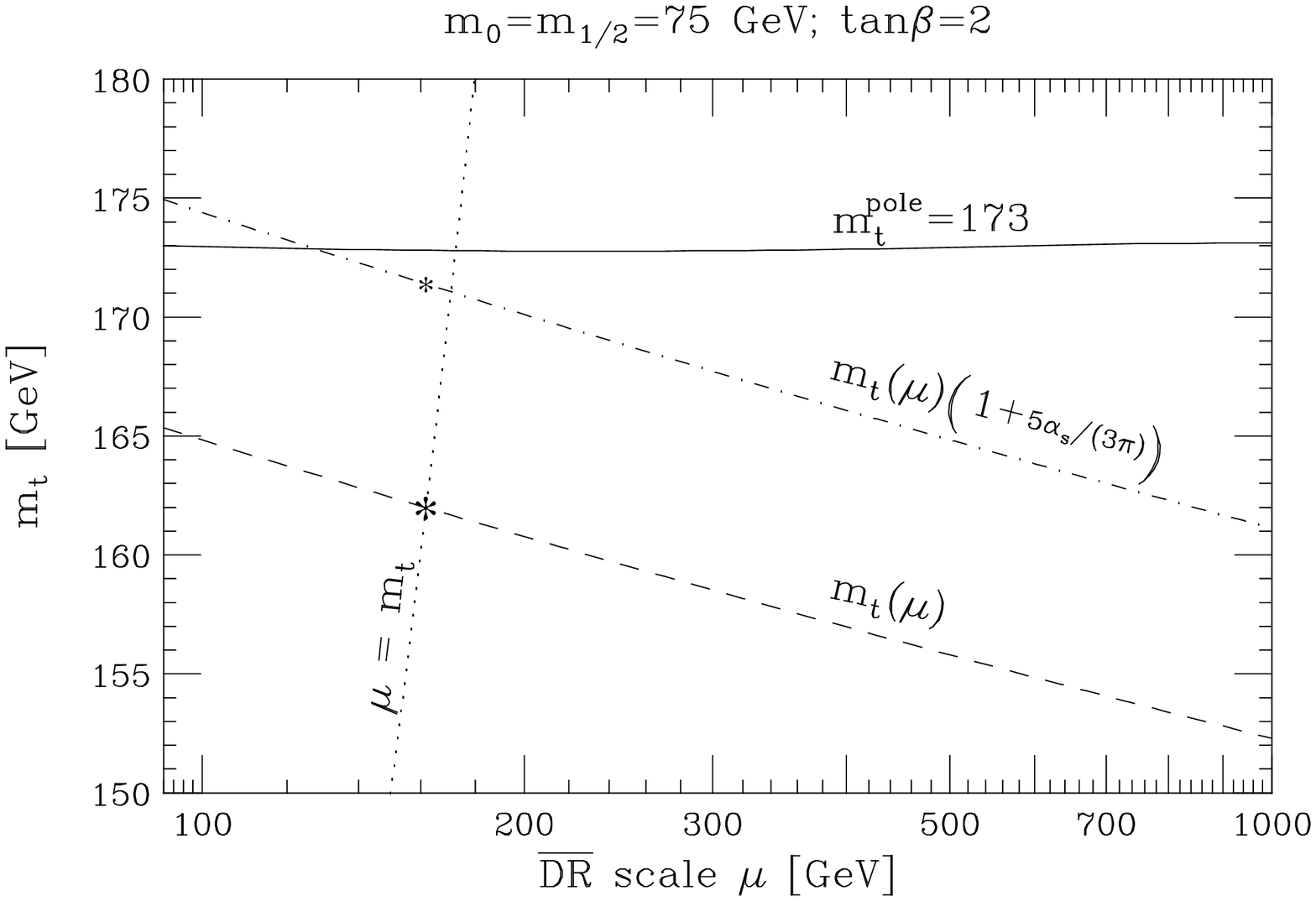}
\parbox{5.5in}{
\caption[]{\small The top quark mass vs. the \dr scale $\mu$. The dashed
line shows the running mass. The dot-dashed line shows the running mass
with the gluon correction taken into account.
The solid line indicates the pole mass, and the horizontal dashed line
shows the points $\mu=m_t$. The asterisks indicate $m_t(m_t)$ before and
after including the gluon correction.
}}
\end{center}
\end{figure}
\dr mass evaluated at the scale $m_t$ yields the approximate
pole mass $m_t(m_t)=162$ GeV. In this example $m_0$ and
$m_{1/2}$ are chosen small so that the squark and gluino masses are of
order $m_t$. The gluino/squark correction in this case is small.
The running mass with only the standard gluon
correction taken into account gives 171 GeV, a satisfactory approximation
to the complete one-loop mass of 173 GeV.

\section{Yukawa coupling corrections}
In this section I explain how to obtain the threshold corrections
to the Yukawa couplings using the mass threshold corrections. Here we
consider the top quark mass $m_t$ as an input. We want to then determine
the \dr Yukawa coupling which runs according to the full undecoupled
MSSM RGE. This
will give us the boundary condition on the Yukawa coupling which we can
then run up to the GUT scale. This \dr Yukawa coupling contains in it all
of the weak scale threshold corrections.

We know the relation between
the top quark pole mass and the running \dr mass
$$m_t^{\rm pole} = m_t^{\ds}\!(\mu) - \Sigma_t(m_t)$$
where $\Sigma_t$ is the top quark self energy\footnote{
$\Sigma_t$ signifies $\Sigma_1 + m_t\Sigma_\gamma$, where the quark self
energy is written $\Sigma_1 + \Sigma_\gamma\rlap/p
+ \Sigma_{\gamma5}\rlap/p\gamma_5 + \Sigma_5\gamma_5$.}, and the running
\dr quark mass is related to the \dr vev and \dr Yukawa
coupling by
$$m_t^{\ds}\!(\mu)={1\over\sqrt{2}}\lambda_t^{\ds}\!(\mu)v_2^{\ds}\!(\mu)\ .$$
Hence we can solve for the threshold-corrected \dr Yukawa coupling if we
know the \dr vev. The \dr vev is determined from the gauge couplings and
the $Z$-boson mass,
$$M_{Z_{\rm pole}}^2 = {1\over4}(g^2 + {g'}^2)(v_1^2 + v_2^2)
- \Pi^T_{ZZ}(M_Z^2)$$
where $g$ and $g'$ are the \dr $SU(2)_L$ and $U(1)_Y$ gauge couplings,
$v_1$ and $v_2$ are the \dr vev's, and $\Pi^T_{ZZ}$ is the transverse
part of the $Z$-boson self energy. Thus we have the threshold
corrected Yukawa coupling,
$$\lambda_t^{\ds}\!(\mu) = \left[{g^2+{g'}^2\over
M_{Z_{\rm pole}}^2 + \Pi^T_{ZZ}}\right]^{1/2}{m_t^{\rm pole} + \Sigma_t\over
\sqrt{2}\sin\beta}\ .$$
This formula gives the threshold-corrected Yukawa coupling in
terms of two point functions, and the two-point function self energy
formulae are much simpler than those obtained by using three point diagrams.
Thus we demonstrate a simple way to go from the measured top quark mass to the
threshold-corrected Yukawa coupling.

\section{Results and summary}

In Fig.6(a) I show the correction to the top quark mass versus
$m_{1/2}$ for $m_0=200$ GeV, $\tan\beta$=3 and $A_0$=0. I indicate the
gauge/Higgs, gluino and gluon contributions separately. The gauge/Higgs
contribution is very small, between $-1$ and 0\% on this plot. The
gluon contribution is constant at 6\%, and the gluino contribution
grows logarithmically with $m_{1/2}$, surpassing the gluon correction
for $m_{1/2}>480$ GeV.
Fig. 6(b) shows the top quark mass correction vs. $m_0$
for $m_{1/2}=200$ GeV, $\tan\beta=2$ and $A_0=0$ GeV. Here the
gauge/Higgs contribution is smaller than $0.4\%$, while the gluino
contribution is an almost constant 3.3\%.
\begin{figure}[ht]
\epsfysize=2.8in
\begin{center}
\epsffile[20 507 610 767]{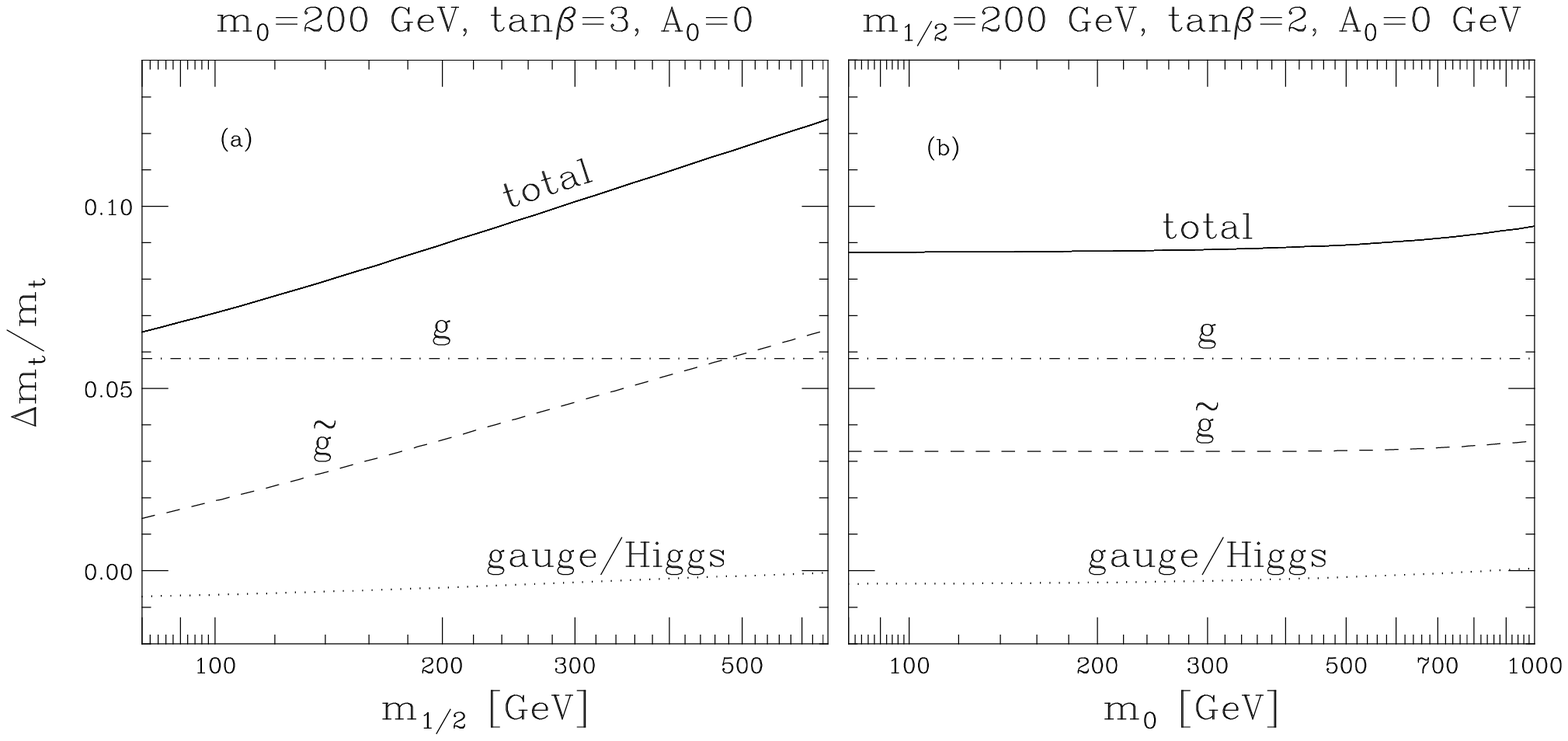}
\parbox{5.5in}{
\caption[]{\small The threshold corrections to the top quark mass
vs. (a) $m_{1/2}$ (b) $m_0$.
The gauge/Higgs, gluino, gluon and total corrections are indicated.
}}
\end{center}
\end{figure}

In Figs. 7(a) and (b)
\begin{figure}[htb]
\epsfysize=2.8in
\begin{center}
\epsffile[25 507 610 757]{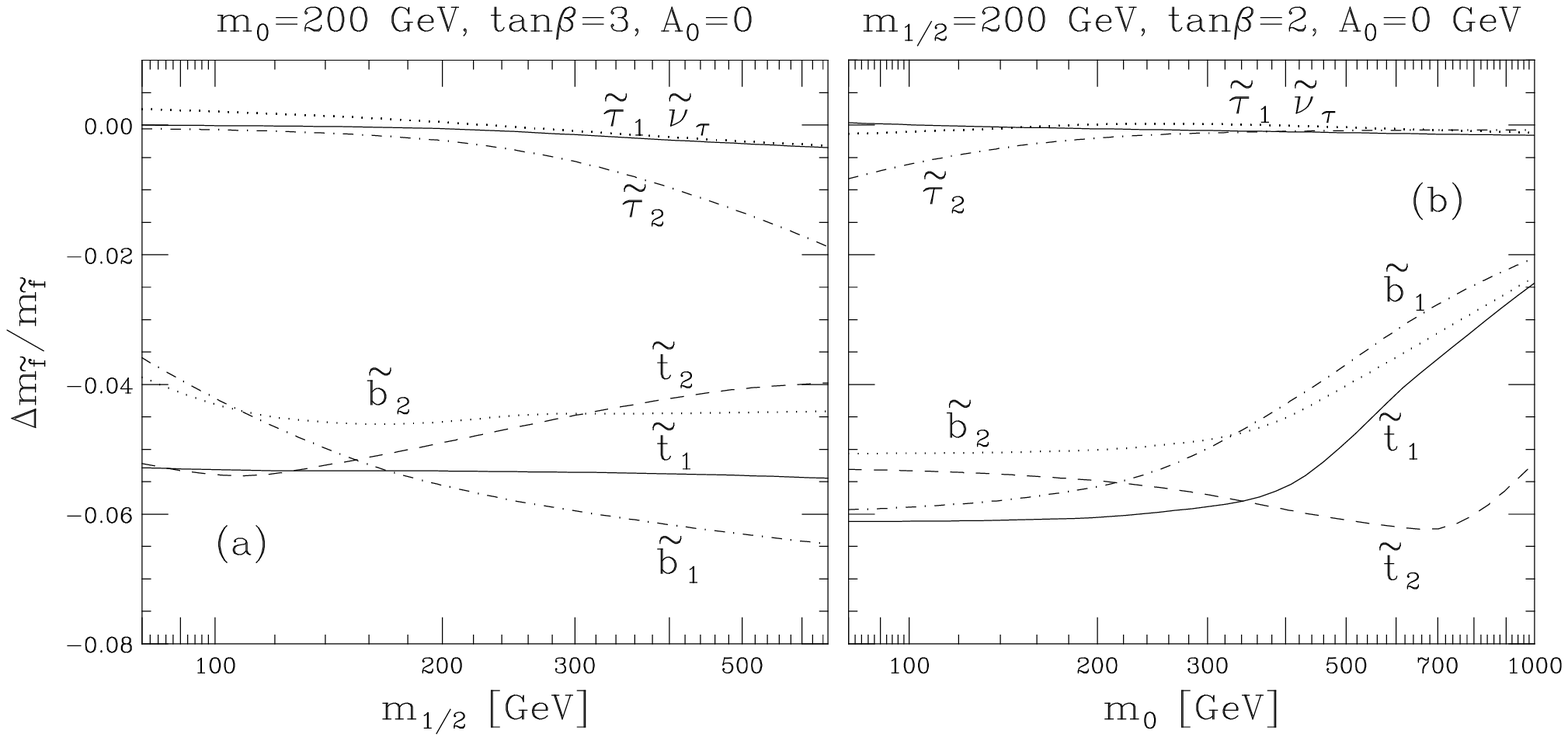}
\parbox{5.5in}{
\caption[]{\small The threshold corrections to sfermion masses vs.
(a) $m_{1/2}$ (b) $m_0$.
}}
\end{center}
\end{figure}
I show the corrections to the third generation squark and slepton
masses for the
same values of parameters as in Figs. 6(a) and (b).
For this choice of parameters the squark mass corrections are negative
and in the range $-2$ to $-7\%$, and the slepton masses receive small
$\roughly{<}1\%$ corrections. For each curve in these plots the
renormalization scale is set equal to the mass.
The quark, squark, lepton and slepton mass
corrections are treated more thoroughly in Ref.\cite{complete}.

In this talk I have emphasized that the leading logarithmic
weak scale threshold corrections to the masses in the MSSM do not
generally dominate the finite corrections. The logarithmic
corrections are taken into account in the ``match and run"
procedure. Hence, when large mass hierarchies are present the ``match and
run" procedure may be useful in approximating the corrections.
In the low energy mass spectra of the MSSM the particle masses are not
generally widely separated and the ``match and run" approximation to
the threshold corrections can lead to errors of order 1. The finite and
logarithmic corrections are taken into account consistently when using the
exact one-loop threshold functions and this leads to precise results.
By using the two-loop RGE's and by taking into account the weak scale
thresholds correctly we can reliably investigate the implications of
GUT scale boundary conditions.

\vskip .3in
{\noindent \Large\bf Acknowledgements}
\vskip .2in
\noindent I would like to acknowledge my collaborators Jonathon Bagger,
Konstantin Matchev and Renjie Zhang. This work was supported by the
NSF under grant PHY-90-9619.

\end{document}